\documentstyle[11pt,twoside,colloqOHP,epsfig]{article}
\begin{document}
\title{Infrared Radiation from Hot Jupiters}
\author{Drake Deming$^1$, L. Jeremy Richardson$^2$, Sara Seager$^3$, 
Joseph Harrington$^4$}
\affil{$^1$Planetary Systems Laboratory, NASA's Goddard Space Flight Center,
 Code 693, Greenbelt, MD 20771 USA [ddeming@pop600.gsfc.nasa.gov]} 
\affil{$^2$NRC Research Associate in the Exoplanet and Stellar Astrophysics 
Laboratory, NASA's Goddard Space Flight Center, Code 667, 
Greenbelt MD 20771 USA 
[richardsonlj@milkyway.gsfc.nasa.gov]} 
\affil{$^3$Department of Terrestrial Magnetism, Carnegie Institution 
of Washington, 5241 Broad Branch Road NW, Washington, DC 20015 USA 
[seager@dtm.ciw.edu]} 
\affil{$^4$Center for Radiophysics and Space Research, Cornell University,
326 Space Sciences Bldg., Ithaca, NY 14853-6801 USA 
[jh@oobleck.astro.cornell.edu]} 

\begin{abstract}
Recent Spitzer infrared (IR) observations of two transiting hot
Jupiters during their secondary eclipses have provided the first direct
detection of planets orbiting other stars (Charbonneau et al. 2005;
Deming et al. 2005).  We here elaborate on some aspects of our
detection of HD\,209458b at 24~$\mu$m, and we compare to the detection
of TrES-1 by Charbonneau et al.  Spitzer will eventually determine the
IR spectral energy distribution of these and similar hot Jupiters,
opening the new field of comparative exoplanetology.  For now, we have
only three Spitzer data points, augmented by upper limits from the
ground.  We here interpret the available measurements from a purely
observational perspective, and we point out that a blackbody spectrum
having $T \sim 1100$K can account for all current IR measurements,
within the errors.  This will surely not remain true for long, since
ongoing Spitzer observations will be very sensitive to the IR
characteristics of hot Jupiters.
\end{abstract}
\section{Direct Detection of Extrasolar Planets}

There are now over 160 extrasolar planets known from Doppler surveys,
and about 15\% of these are hot Jupiters.  Strong stellar irradiation
heats these planets to $T > 1000$K (Seager and Sasselov 1998), so they
should emit most of their radiation in the IR.  A number of pioneering
attempts were made to detect hot Jupiters from the ground in the
combined IR light of star+planet (Wiedemann et al. 2001, Lucas and
Roche 2002, Richardson et al. 2003a,b). But success was only recently
achieved, and it required using the Spitzer Space Telescope
(Charbonneau et al. 2005; Deming et al. 2005).

The Spitzer observations of two hot Jupiters are the first direct
detections of extrasolar planets.  To be clear about terminology, by
direct detection we mean that photons emitted by the planet are
detected, and are separated from stellar photons by some method.  One
method to separate the stellar and planetary photons would be to
spatially resolve the planet and the star, e.g., by imaging.  Imaging
of extrasolar planets is being widely pursued, but it requires a very
high order of technology. The technique which is successful for
Spitzer is to observe the {\it secondary eclipse} of transiting hot
Jupiters.  We measure the star+planet when the planet is out of
eclipse, and the star alone when the planet is eclipsed, and the
diference tells us how many photons are due to the planet.

Note that the secondary eclipse technique does not specify {\it which}
of the detected photons are due to the planet, and which are from the
star, only {\it how many} photons are from the planet.  But for
scientific purposes, `how many' is the primary quantity of interest!
So secondary eclipses are a very sensitive and useful method for
direct detection and characterization of transiting extrasolar
planets.  There are {\it nine} ongoing Spitzer programs to detect and
characterize close-in extrasolar planets in combined IR light, and we
are now truly in the era of comparative exoplanetology.

In Sec 2, we describe our detection of the secondary eclipse of
HD\,209458b, elaborating on some points not mentioned in our detection
paper.  Sec. 3 interprets our results in combination with the TrES-1
measurements by Charbonneau et al. (2005), and factors in the
ground-based upper limits.  We point out the importance of measuring
these planets in the 2- to 5-$\mu$m wavelength region, where there is
currently a considerable question about their level of IR emission.

\section{The Secondary Eclipse of HD\,209458b}

Figure 1 shows the secondary eclipse of HD\,209458b at 24 $\mu$m
wavelength, observed using the Multiband Imaging Photometer for
Spitzer (MIPS, Rieke et al. 2004).  The scatter in the individual
points (upper panel) is about 0.008 magnitudes, poorer than is
possible from the ground in the visible.  However, the MIPS precision
is limited by statistical fluctuations in the thermal emission from
dust in our own solar system, the IR zodiacal light.  The Spitzer data
analysis pipeline provides an `error image' which quantifies these
statistical fluctuations.  We extract the brightness of the star using
an optimal weighting technique (Horne 1986), and we propagate the
error image thru the same procedure to derive the formal errors in the
photometry.  We find the scatter in our photometry to be in close
agreement with the formal errors.  We are background-limited by the
thermal zodiacal emission.

\begin{figure}
\epsfig{file=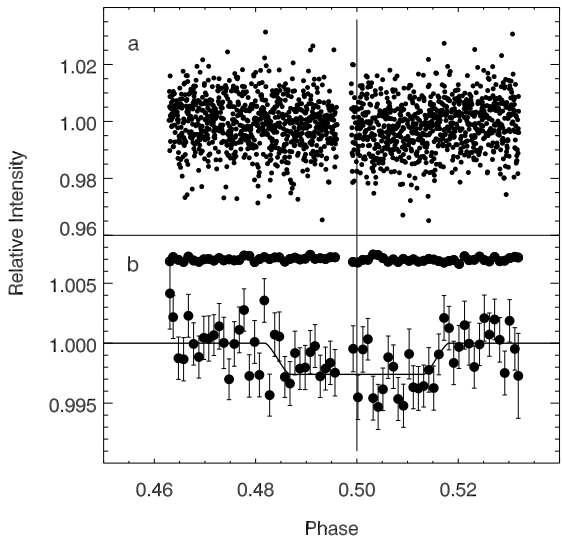, width=8cm}
\caption{Photometric detection of the secondary eclipse of 
HD\,209458b using Spitzer/MIPS.  The upper panel shows all 1696
individual photometric points, which are binned in the lower panel to
show the eclipse.  The fitted eclipse curve is overplotted.  The
points above the eclipse are a control sequence, showing the stability
of the technique.  Note that the eclipse is accurately centered at
phase 0.5, as marked by the vertical line.}
\end{figure}

In addition to noise from the zodiacal background, there are some
instrument quirks in MIPS.  For example, there is a `first frame'
effect, where the signal from the MIPS detector is different if it has
just been reset.  Fortunately, we were able to use a simple trick to
remove the instrument quirks.  Our trick is to ratio the intensity of
the star to the total intensity of the zodiacal background in the
image.  Since the background is a large signal, its relative precision
is much better than the star's precision, and the ratio does not add
significant additional noise. Moreover, because of Spitzer's modest
(0.85-meter) aperture, diffraction spreads the stellar image over
multiple pixels, like the background.  Also, the per-pixel intensity
of the star is only modestly greater than the background.  To MIPS,
the star looks like a small patch of bright background, so the
instrument treats the background and stellar photons exactly the same
- it cannot `tell the difference'.  After normalizing the stellar
intensity to the background, we find the instrument quirks are gone,
and the noise is accurately characterized as Gaussian white noise,
which averages down as the square-root of the number of observations -
just as the text books predict.

The lower panel of Figure 1 averages our secondary eclipse photometry
into bins of 0.001 in phase, and adds error bars to the binned values.
Now the secondary eclipse is quite evident, and is seen to be centered
at phase 0.5.  We fit an eclipse curve to the data, and find a
best-fit depth of $0.26\pm0.046\%$, a $5.6\sigma$ detection. The
planet's brightness temperature is $1130\pm150$K.  Simultaneously with
our MIPS detection of HD\,209458b, Charbonneau et al. (2005) detected
TrES-1 at 8- and 4.5-$\mu$m using Spitzer's InfraRed Array Camera
(IRAC).  The TrES-1 measurements imply a very similar temperature
($1060$K), and their 8 $\mu$m measurment has over $6\sigma$ statistical
significance.  Since these three data points are (so far) our only direct
measurements of hot Jupiters, they have elicited considerable
interest.  Four modeling papers have interpreted the measurements
(Barman et al. 2005, Burrows et al. 2005, Fortney et al. 2005, Seager
et al. 2005), but in Sec. 3 we will give a more simple-minded
interpretation, highlighting what we call the `short IR wavelength
question'.

The Spitzer detections have been described as surprising, in the sense
of being unexpected.  Certainly this application for Spitzer was
unanticipated in the early years when the observatory was being
developed.  But recently, the planet-to-star contrast ratio was
robustly predicted to be a significant fraction of one percent at long
IR wavelengths (Charbonneau 2003, Burrows et al. 2004).  Prior to
launch, we reasoned that if we couldn't detect a fraction of one
percent using a cryogenic telescope in space, then we should surely
give up!  Fortunately, the first public data from MIPS showed us that
Spitzer's sensitivity and stability were up to the task.  So we knew
in advance that we would detect HD\,209458b with MIPS, providing that
the planet was at least as hot as 700K - which seemed unavoidable.

Curiously, the Spitzer detections did not start with the best cases.
Since the flux from both the planet and star decrease with wavelength
(decline of the Planck function), 24~$\mu$m is not the best wavelength
for secondary eclipse detection.  The IRAC 8~$\mu$m channel is more
suitable, being source photon-limited, not background-limited.  But
for historical reasons, IRAC was first used by Charbonneau et al. for
TrES-1, not for HD\,209458.  So the brighter and closer system was not
observed at the best wavelength.  The new observing programs now
underway will soon fill this gap, and we will see secondary eclipses
of hot Jupiters even more clearly.  The new very hot Jupiter recently
announced by Bouchy et al. (2005) will be an especially important
target for Spitzer.

\section{The Short IR Wavelength Question}

\subsection{New 3.8 $\mu$m Photometry}

In order to discuss what we regard as the major unresolved
observational question concerning hot Jupiters, we need to include the
ground-based data. A flux peak is predicted to occur in hot Jupiter
spectra at 3.8 $\mu$m (Sudarsky et al. 2003).  Ground-based
observations can be very useful at this important wavelength, because
we can use a filter centered exactly on the predicted peak (the IRAC
bandpasses are offset).  Unfortunately, the high thermal background
from the ground has made broadband photometry impossible near 4~$\mu$m
- most instrument detector arrays saturate in quite short integration
times.

\begin{figure}
\epsfig{file=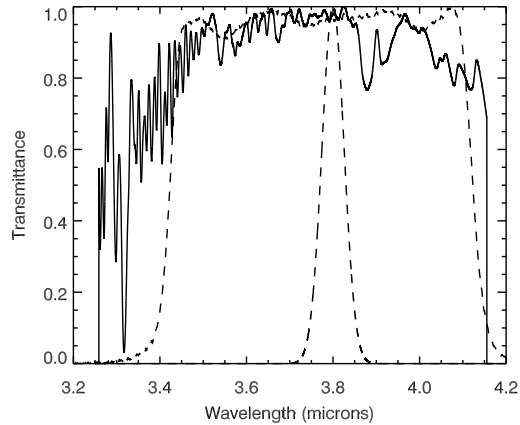, width=8cm}
\caption{Window in the terrestrial atmosphere for photometry of hot 
Jupiters at 3.8 $\mu$m.  The solid line shows the telluric
transmittance at moderate spectral resolution.  The dashed curves are
the transmittance of the conventional L$^\prime$ filter (broad), and our
CVF (narrow).}
\end{figure}

In September 2003 we obtained 3.8 $\mu$m photometry of HD\,209458
during two secondary eclipses, using NSFCAM (Shure et al. 1994) on the
NASA Infrared Telescope Facility (IRTF).  We were able to avoid
detector saturation by using a narrow optical bandwidth, a $1.5\%$
circular-variable-filter (CVF).  Figure 2 shows the atmospheric
transmittance, and the profile of the conventional L$^\prime$ filter, and
our CVF at 3.8 $\mu$m.  To monitor changes in atmospheric absorption,
we observed a comparison star of similar brightness to HD\,209458. 
We have completed the analysis of one of the two eclipses.  The
eclipse amplitude in that case, from 329 individual 10-second
exposures, was $-0.0007\pm0.0014$, i.e. the system nominally becomes
brighter when the planet is hidden, consistent with seeing no eclipse.
Not surprisingly, the errors are larger in the ground-based
observations than with Spitzer, but nevertheless the results are
sufficiently precise to be somewhat puzzling, as we explain below.

\subsection{An Observational Perspective}

The two extrasolar planets currently observed by Spitzer - HD\,209458b
and TrES-1 - are different worlds, and they surely have their own
unique characteristics.  But, our ignorance of hot Jupiter spectra is
arguably much greater than the real differences between them. So here 
we compare observations of both planets to a single model.  To
quantify the similarity between the two planets, we assume thermal
equilibrium and the same Bond albedos and efficiency of heat
redistribution.  In the long wavelength (Rayleigh-Jeans) limit, it is
easy to show that the planet-to-star contrast ${c_\lambda}$ depends on
the stellar and orbital parameters as:
\begin{equation}
{c_\lambda} \sim  {R_p^2}{R_*^{-{\frac{3}{2}}}}{a^{-{\frac{1}{2}}}}
\end{equation} 
where $R_p$ and $R_*$ are the planetary and stellar radius, 
respectively, $a$ is the orbit semi-major axis, and we supress the
constant containing the Bond albedo.  Note that the temperature of the
star does not appear since it cancels when computing the
contrast. Using the parameters for HD\,209458 (Brown et al. 2001;
Wittenmyer 2005) and TrES-1 (Alonso et al. 2004), we find that the contrasts
for these two systems at a given wavelength would differ by only 8\%
under our simple assumptions.  So comparing them to a single model is
reasonable.

Figure 3 shows contrast versus wavelength for HD\,209458b and TrES-1,
and compares the observations to a model from Sudarsky et al. (2003),
as did Charbonneau et al. (2005, their Figure 3).  The dashed line is
the contrast from a blackody having $T=1100$K.  The Spitzer
observations cannot in themselves discriminate between the model curve
and a simple blackbody.  Considering the errors, all three Spitzer
observations could be consistent with either the Sudarsky et al. model
or the blackbody.

\begin{figure}
\epsfig{file=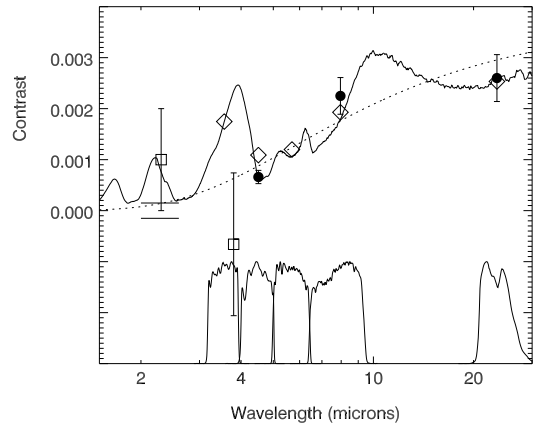, width=8cm}
\caption{Spitzer and ground-based observations of TrES-1 and
HD\,209458b in contrast units, compared to a single model (solid line)
and 1100K blackbody (dotted line).  The Spitzer IRAC and MIPS
bandpass funtions are plotted below.  Filled symbols are the Spitzer
points, and the open diamonds show the expectation value of the
contrast averaged over the Spitzer bands. The open squares show the
ground-based photometry of Snellen (2005) at 2.3~$\mu$m, and from this
paper at 3.8~$\mu$m.  The two horizontal lines at 2.2~$\mu$m represent
the limit from Richardson et al. (2003b).  Note log scale for
wavelength.}
\end{figure}

Now we consider the ground-based data.  The error bar on our CVF
photometry at 3.8 $\mu$m misses the Sudarsky et al. peak at this
wavelength, but it overlaps the blackbody.  The Snellen (2005)
photometry at 2.3 $\mu$m agrees with the model, but the error bar also
includes the blackbody curve.  The most stringent ground-based point
is the Richarson et al. (2003b) upper limit at 2.2 $\mu$m.  This is a
limit, derived from differential spectroscopy, on the shape of the
spectrum, and it is given by the two horizontal lines.  Unlike the
other error limits on the figure, this is a 3$\sigma$, not 1$\sigma$
limit.  These lines are relative limits, i.e. only the intensity
interval between them is significant (see Seager et al. 2005). To
conform to this limit, the contrast spectrum has to fit within the
lines. The Sudarsky et al. model doesn't fit, but the blackbody does.
This is the essence of the short-IR wavelength question - we don't see
the peaks predicted in the spectrum where planet flux should escape
between absorption bands, and we wonder how strong these peaks really
are.  Now, we do not seriously suggest that the planet is actually a
blackbody.  But to date, we do not have a definitive measurement of
departures from a blackbody spectrum in the IR.  The IRAC band at 3.5
$\mu$m should provide this, as evident from Figure 3.  Observations of
TrES-1 at 3.5 $\mu$m were made by Spitzer in September 2005, and are
now being analyzed by Dave Charbonneau.  We eagerly await the results.


\end{document}